\documentclass[conference]{IEEEtran}
\usepackage{latexsym}
\usepackage{amsfonts}
\usepackage{amssymb}
\usepackage{amsbsy}
\usepackage{amsmath}
\usepackage{amsthm}
\usepackage{color}
\usepackage{pgf}
\usepackage{graphicx,lscape,rotating,subfig}
\usepackage{balance}
\usepackage[varg]{txfonts}
\usepackage{algorithm,algpseudocode}
\usepackage[noadjust]{cite}

\theoremstyle{remark}

\theoremstyle{definition}

\newcommand{\CoMPtwoflex}{\mathsf{CoMP2flex}}
\newcommand{\CoMP}{\mathsf{CoMP}}
\newcommand{\CoMPflex}{\mathsf{CoMPflex}}

\newcommand{\Ebb}{\mathbb{E}}
\newcommand{\lt}{\mathcal{L}}
\def\({\left(}
\def\){\right)}

\title{On a User-Centric Base Station Cooperation Scheme for Reliable Communications}

\author{\IEEEauthorblockN{Dong Min Kim, Henning Thomsen and Petar Popovski}
\IEEEauthorblockA{Department of Electronic Systems, Aalborg University, Denmark\\
Email: \{dmk;ht;petarp\}@es.aau.dk} }

\begin{document}
\maketitle

\begin{abstract}
In this paper, we describe $\CoMPtwoflex$, a user-centric base station (BS) cooperation
scheme that provides improvements in reliability of both uplink (UL) and downlink (DL)
communications of wireless cellular networks. $\CoMPtwoflex$ supports not only
cooperation of two BSs with same direction of traffic but also cooperation of two BSs
serving bidirectional traffic. The reliability performance of $\CoMPtwoflex$ is shown
with numerical simulations and analytical expressions. We quantify and numerically
validate the performance of the greedy BS pairing algorithm by comparing maximum weight
matching methods, implemented as the Edmonds matching algorithm for weighted graphs.
\end{abstract}


\section{Introduction}

Enhancing the reliability of communication, even enduring low data rate, is one of the 5G
wireless system design factors \cite{Popovski2014ultra}. Previously, many researchers
have focused on the downlink (DL) traffic due to its greater volume compared to the
uplink (UL) traffic. Recently, UL traffic has increased because of new mobile
applications and the massive growth of Internet-of-Things (IoT) applications. Many IoT
applications have an intensive UL traffic by their sensing and monitoring characteristics
\cite{shafiq2013large}. For this reason, enhancing the reliability of UL transmission as
important as that of DL transmission.

Increasing the number of base stations (BSs) per area, \emph{network densification}, is
one way to increase the reliability. By taking advantage of multiple proximate and
interconnected BSs, a BS cooperation in same transmission direction is considered in
terms of coordinated multipoint transmission ($\CoMP$)
\cite{karakayali2006network,sawahashi2010coordinated,lee2012coordinated,baccelli2015stochastic}.
$\CoMP$ can increase the reliability performance, however, $\CoMP$ considers cooperation
between the BSs with the same direction of traffic.

In \cite{thomsen2015compflex,thomsen2016full}, a cooperation scheme for serving cross
directional (UL and DL) traffic simultaneously utilizing separated half duplex BSs is
proposed and investigated, and is termed $\CoMPflex$: $\CoMP$ for In-Band Wireless
Full-Duplex. The limitation of \cite{thomsen2015compflex,thomsen2016full} is that the
traffic is assumed always cross directional. In general, the cooperative BSs could serve
the same or opposite direction of traffic. This means that the reliability might be
improved if the network can support both $\CoMP$ and $\CoMPflex$ according to the
traffic. In this paper, we propose $\CoMPtwoflex$, a user-centric base station (BS)
cooperation scheme that provides improvements in reliability of both uplink (UL) and
downlink (DL) communications of wireless cellular networks. The proposed $\CoMPtwoflex$
can support not only cooperation of two BSs with same direction of traffic but also
cooperation of two BSs serving bidirectional traffic. We will show that the reliability
performance of $\CoMPtwoflex$ by analytical expressions and numerical simulations. We
quantify and numerically validate the performance and complexity of the greedy BS pairing
algorithm by comparing the Edmonds matching algorithm for weighted graphs, one of maximum
weight matching methods.

The paper is organized as follows. In Section~II, we describe the system model of the
proposed scheme. The reliability analysis of proposed scheme is given in Section~III, and
its numerical results are presented in Section~IV. The paper is concluded in Section~V.

\section{System Model of $\CoMPtwoflex$}

\begin{figure}[tb]
\centering
\subfloat[Dynamic Cell Selection for DL-$\CoMP$]{
\includegraphics[width=0.32\linewidth]{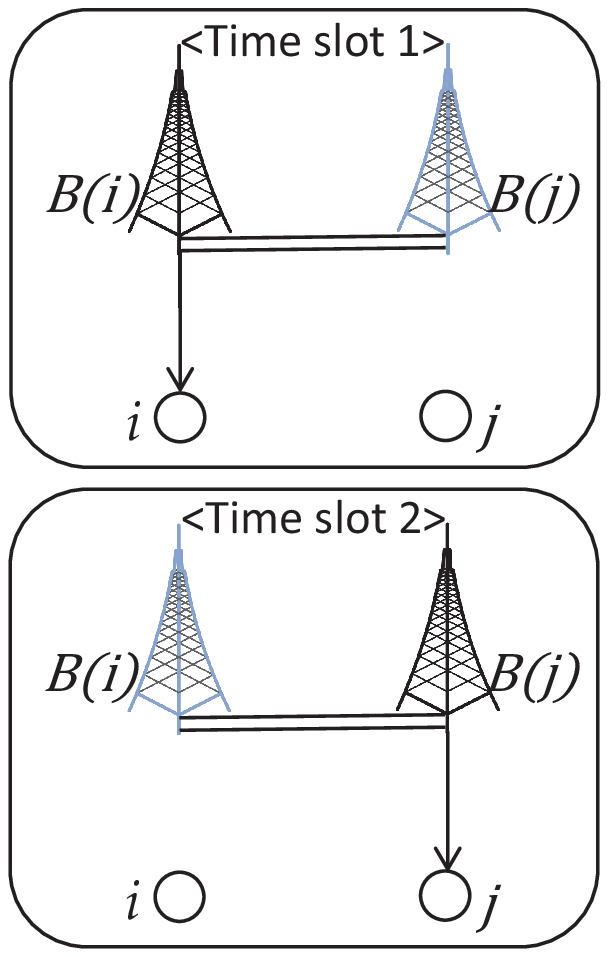}
\label{F:DLCoMP}}
\subfloat[Coordinated Scheduling for UL-$\CoMP$]{
\includegraphics[width=0.32\linewidth]{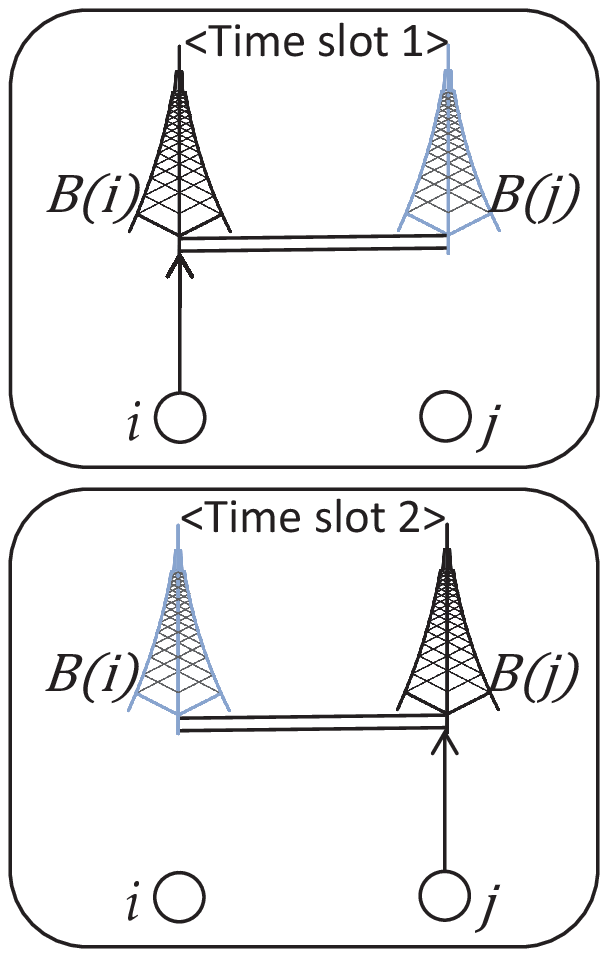}
\label{F:ULCoMP}}
\subfloat[$\CoMPflex$]{
\includegraphics[width=0.32\linewidth]{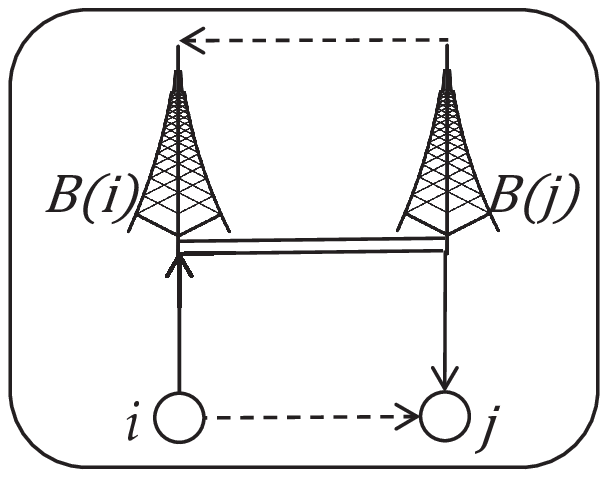}
\label{F:CoMPflex}}
\caption{Different BS cooperation modes of $\CoMPtwoflex$.}
\label{F:comp2flex}
\end{figure}

We propose a combined $\CoMP$ and $\CoMPflex$ ($\CoMPtwoflex$) BS cooperation scheme,
which can support not only cooperation of two BSs with same direction of traffic but also
cooperation when the BSs serve opposite directions of traffic. We consider a pair of
cooperating BSs. They are interconnected via a wired connection (double solid line in
Fig.~\ref{F:comp2flex}). If both BSs serve DL traffic, dynamic cell selection (DCS) is
used (Fig.~\ref{F:DLCoMP}). DCS serves one user in a cell first, and then serves the user
in the other cell. If both BSs serve UL traffic, $\CoMP$ reception with coordinated
scheduling is used, i.e., only one user in one cell will transmit and the user in the
other cell will transmit later (Fig.~\ref{F:ULCoMP}). Lastly, if the
 BSs serve cross directional traffic, one BS will operate in DL (DL-BS) and
the other in UL (UL-BS), or vice-versa (Fig.~\ref{F:CoMPflex}). The UL-BS uses side
information sent from the DL-BS through the wired backhaul, for interference
cancellation.

To quantify the performance of $\CoMPtwoflex$, we assume that the BSs are randomly
distributed with density $\lambda_\mathrm{B}$, resulting in a homogeneous Poisson point
process (PPP) \cite{Chiu2013stochastic}. The mobile stations (MSs) are associated with
the nearest BS, and each BS serves one MS using the same resource at a time. We assume
that each MS, independently of one another, selects its transmission direction with a
certain probability $\delta$. We can define a random variable $T$ for the traffic
direction, $\Pr\{T=\mathrm{DL}\}=\delta$ and $\Pr\{T=\mathrm{UL}\}=1-\delta$. The traffic
asymmetry of the network can be adjusted by changing $\delta$. A single frequency and
Rayleigh fading channel with unit mean power is assumed. $\ell(r)=r^{-\alpha}$ is the
path-loss function with path loss exponent $\alpha$. The default transmit power of uplink
users is $P_\mathrm{M}$. BSs transmit with constant power $P_\mathrm{B}$. A downlink
(uplink) signal-to-interference-plus-noise ratio (SINR) requirement is denoted by
$\beta_\mathrm{D}$ ($\beta_\mathrm{U}$). The term $\sigma^2$ is additive white Gaussian
noise. The transmission is successfully received if the received SINR is greater than or
equal to the target threshold value. Each BS can cooperate with an adjacent BS paired by
an algorithm explained in the next subsection.

\subsection{BS Pairing Algorithm}
\label{S:BS_Pairing_Algorithm}

For pairing the BSs, we use two different methods. The first method is a greedy pairing
algorithm, which operates as follows: Given a BS deployment, the algorithm starts with a
BS chosen uniformly at random. It then lists all neighboring BSs and selects the closest
one, in terms of Euclidean distance. These two BSs are then considered a \emph{pair}. The
algorithm then select the next unpaired BS, lists its unpaired neighbors, and pairs it
with the nearest one. When there are no more BSs that can be paired, the algorithm
terminates. Assuming there are $n$ BSs, an upper bound on the complexity is
$n(n-1)$,which is of order $O(n^2)$. To be simple, we restrict the candidate BSs to the
BSs sharing the same edge, so called Delaunay neighbors. Each BS only needs the knowledge
about its Delaunay neighbors, i.e., it is a local method.

The second method is the Edmonds matching algorithm~\cite{galil1986efficient}. This
algorithm needs the knowledge of the entire network, i.e., it is a global method. In this
method, the network is modeled as a weighted planar graph $G = (V,E)$, where the vertex
set $V$ contains the BSs, and $E$ is the edge set, representing connections between the
BSs. There is an edge $e \in E$ if and only if the BSs represented by the endpoints of
$e$ are adjacent in the deployment. The \emph{weight} of an edge $e$, $wt(e)$, is the
distance between the BSs. The goal of the Edmonds matching algorithm is to find a maximal
matching (i.e., containing as many edges as possible), while minimizing the total weight.
Note that in this case, the distance between the typical BS and its paired BS might be
quite high, since the algorithm looks at the entire network.

Our motivations for studying this algorithm are twofold: First, for planar graphs, there
exist implementations with complexity $O(n^{\omega / 2})$, where $\omega < 2.38$ is the
exponent of the best matrix multiplication algorithm~\cite{mucha2006maximum}. Therefore,
the exponent satisfies $\omega/2 < 2.38 / 2 = 1.19$, which is lower than in the greedy
method. Second, we will use the Edmonds matching algorithm to validate the greedy pairing
algorithm.

Based on the system model presented in this section, we will derive the analytical forms
of the transmission success probabilities in UL and DL representing reliability in the
next section.

\section{Reliability Analysis}

In this section, we investigate the reliability performance of $\CoMPtwoflex$ by deriving
analytical expressions for the transmission success probabilities in UL and DL. For the
analytical tractability, we assume that all BSs in the network can schedule the one
\emph{active} mobile station (MS) either UL (UL-MS) or DL (DL-MS) in each Voronoi cell.
We further assume that the spatial distribution of MSs follows the another independent
PPP with the density $\lambda_\mathrm{B}$.

\subsection{UL Success Probability in $\CoMPtwoflex$ Network}

The success probability of transmission of a typical UL user $\mathcal{U}$ at a typical
BS $B\(\mathcal{U}\)$ in $\CoMPtwoflex$ $p_\mathrm{U}^\CoMPtwoflex$ can be expressed as
follows:
\begin{align}\label{E:sucprobULCoMP2flex1}
p_\mathrm{U}^{\CoMPtwoflex} &= \Ebb\left[ \Pr \left[ \frac{g_{\mathcal{U},B\(\mathcal{U}\)}\ell\(r\)P_\mathrm{M}}{I_{B\(\mathcal{U}\)}^\psi+I_{B\(\mathcal{U}\)}^\varphi} \geq \beta_\mathrm{U} \right] \right] \\
    &= \Ebb\left[ \Pr \left[ g_{\mathcal{U},B\(\mathcal{U}\)} \geq \frac{\beta_\mathrm{U} r^\alpha}{P_\mathrm{M}} \(  I_{B\(\mathcal{U}\)}^\psi+I_{B\(\mathcal{U}\)}^\varphi \) \right] \right], \nonumber
\end{align}
where $g_{i,j}$ denotes the gain of the channel at $j$ from $i$, and $I_{i}^\psi$ and
$I_{i}^\varphi$ denote the aggregate interference at node $i$ from DL-BSs and UL-MSs,
respectively. The transmission distance between the typical user and the typical BS is
denoted by $r$, and $r$ is random variable with pdf as $f\(r\)=2\pi \lambda_\mathrm{B} r
\exp\(-\pi \lambda_\mathrm{B} r^2\)$ \cite{andrews2011tractable}. By using the fact that
$g_{\mathcal{U},B\(\mathcal{U}\)}$ is exponential random variable with unit mean (due to
the Rayleigh faded channel), \eqref{E:sucprobULCoMP2flex1} can be expressed as:
\begin{align}\label{E:sucprobULCoMP2flex2}
p_\mathrm{U}^{\CoMPtwoflex}
    &=\Ebb_r\left[
            \Ebb_{I_{B\(\mathcal{U}\)}^\psi,I_{B\(\mathcal{U}\)}^\varphi}\left[
                \exp\( -s \( I_{B\(\mathcal{U}\)}^\psi + I_{B\(\mathcal{U}\)}^\varphi \) \)\right]
          \right] \nonumber \\
    &=\Ebb_r\left[
            \Ebb_{I_{B\(\mathcal{U}\)}^\psi}\left[
                \exp\(-sI_{B\(\mathcal{U}\)}^\psi\)\right]
            \Ebb_{I_{B\(\mathcal{U}\)}^\varphi}\left[
                \exp\(-sI_{B\(\mathcal{U}\)}^\varphi\)\right]
          \right] \nonumber \\
          &\approx \int_0^\infty   \lt_\mathrm{U}^\psi \left( s \right) \lt_\mathrm{U}^\varphi \left( s \right) 2\pi {\lambda _{\mathrm{B}}}r\exp \left( { - \pi {\lambda _{\mathrm{B}}}{r^2}} \right)dr,
\end{align}
where $s = \frac{\beta_\mathrm{U}r^\alpha}{P_\mathrm{M}}$, and $\lt_\mathrm{U}^\psi
\left( s \right)$ and $\lt_\mathrm{U}^\varphi \left( s \right)$ are the Laplace
functionals of the interference from DL-BSs at the typical BS and the interference from
UL-MSs at the typical BS, respectively. The interference from DL-BS is coming from
outside of the pair. Assuming independence of the channels from different interfering
DL-BSs and independence of the distances from different interfering DL-BSs, and using
moment generating function of exponential distribution, $\lt_\mathrm{U}^\psi \left( s
\right)$ can be expressed as:
\begin{align}
\label{E:ltpsiul1}
\lt_\mathrm{U}^\psi \left( s \right)
    &= \Ebb_{I_{B\(\mathcal{U}\)}^\psi}\left[ { \exp\(-\frac{{\beta _{\mathrm{U}}}{r^\alpha}}{P_\mathrm{M}} I_{B\(\mathcal{U}\)}^\psi \) } \right] \nonumber \\
    &= \Ebb_{g_{i,B\(\mathcal{U}\)},r_{i,B\(\mathcal{U}\)}}\left[ { \exp\(-\frac{{\beta _{\mathrm{U}}}{r^\alpha}}{P_\mathrm{M}} \sum\limits_{i \in \Phi^\psi}{g_{i,B\(\mathcal{U}\)}} r_{i,B\left( \mathcal{U} \right)}^{ - \alpha }P_\mathrm{B} \) } \right] \nonumber \\
    &= \Ebb_{r_{i,B\(\mathcal{U}\)}}\left[ {\prod\limits_{i \in \Phi^\psi} \Ebb_{g_{i,B\(\mathcal{U}\)}} \left[\exp\(-\frac{P_\mathrm{B}}{P_\mathrm{M}} {g_{i,B\(\mathcal{U}\)}} {\beta _{\mathrm{U}}}{r^\alpha }r_{i,B\left( \mathcal{U} \right)}^{ - \alpha }\)\right] } \right] \nonumber \\
    &= \Ebb_{r_{i,B\(\mathcal{U}\)}}\left[ {\prod\limits_{i \in \Phi^\psi} {\frac{1}{{1 + \left( {{P_{\mathrm{B}}}/{P_{\mathrm{M}}}} \right){\beta _{\mathrm{U}}}{r^\alpha }r_{i,B\left( \mathcal{U} \right)}^{ - \alpha }}}} } \right],
\end{align}
where $r_{i,B\(\mathcal{U}\)}$ denotes the distance from $i$th BS in the interfering BS
set $\Phi^\psi$ to the typical BS $B\(\mathcal{U}\)$. We assume that the set $\Phi^\psi$
follows PPP with density $\lambda_\mathrm{B}^\psi$. To approximate the density
$\lambda_\mathrm{B}^\psi$, we further assume that entire DL-BSs have cooperation BSs. The
average node density of DL-BSs is $\delta\lambda_\mathrm{B}$. DL-BSs are either $\CoMP$
BSs or $\CoMPflex$ BSs. A pair of BSs can be $\CoMP$ BSs when both BSs serve DL traffic,
hence the node density of $\CoMP$ BSs is $\delta^2\lambda_\mathrm{B}$. The rest of DL-BSs
are $\CoMPflex$ BSs, hence their node density is $\delta\(1-\delta\)\lambda_\mathrm{B}$.
Only half of $\CoMP$ BSs can be simultaneously active due to adopting DCS mode. So,
$\lambda_\mathrm{B}^\psi$ is equal to
$0.5\delta^2\lambda_\mathrm{B}+\delta\(1-\delta\)\lambda_\mathrm{B}=\(\delta-0.5\delta^2\)\lambda_\mathrm{B}$.
In case of the typical BS having the DL-BS as pair, the interference from the paired
DL-BS to the typical UL-BS is cancelled. Hence the interference at the UL-BS is coming
from DL-BSs not paired to it. Assuming the paired BS is the nearest DL-BS, the distance
to the nearest \emph{interfering} BS is the distance to the second nearest DL-BS. The
second nearest distance distribution is given as \cite{moltchanov2012distance}:
\begin{align}\label{E:2ndnearest}
f\(d\)=2\(\pi\lambda\)^2 d^3 \exp\(-\pi\lambda d^2\).
\end{align}
Using \eqref{E:2ndnearest}, $\lambda_\mathrm{B}^\psi$, and probability generating
functional (PGFL) of PPP \cite{Chiu2013stochastic}, \eqref{E:ltpsiul1} can be expressed
as:
\begin{align}
\label{E:ltpsiul2}
\lt_\mathrm{U}^\psi \left( s \right) &= \int_0^\infty
    \exp \left( { - 2\pi \lambda _{\mathrm{B}}^\psi
            \int_t^\infty  {\frac{{\left( {{P_{\mathrm{B}}}/{P_{\mathrm{M}}}} \right){\beta _{\mathrm{U}}}{r^\alpha }{x^{ - \alpha }}}}{{1 + \left( {{P_{\mathrm{B}}}/{P_{\mathrm{M}}}} \right){\beta _{\mathrm{U}}}{r^\alpha }{x^{ - \alpha }}}}xdx} }
         \right)\cdot \nonumber \\
&\qquad\qquad 2 \(\pi \lambda _{\mathrm{B}}\)^2 t^3 \exp \left( { - \pi \lambda _{\mathrm{B}} {t^2}} \right)dt,
\end{align}
where $t$ is the distance to the nearest interfering DL-BS (second nearest BS). For
brevity, $r_{i,B\(\mathcal{U}\)}$ is changed as $x$.

In a similar way, we can obtain $\lt_\mathrm{U}^\varphi \left( s \right)$. The
interfering MS set will be denoted $\Phi^\varphi$ and it is assumed as PPP with density
$\lambda_\mathrm{B}^\varphi$. To approximate the density $\lambda_\mathrm{B}^\varphi$, we
assume that all UL-BSs serving UL-MSs are paired. The average node density of UL-MSs is
$\(1-\delta\)\lambda_\mathrm{B}$. A pair of UL-MSs can be $\CoMP$ MSs when both MSs have
UL traffic, hence the node density of $\CoMP$ MSs is $\(1-\delta\)^2\lambda_\mathrm{B}$.
The rest of UL-MSs are $\CoMPflex$ MSs, hence it is
$\(1-\delta\)\delta\lambda_\mathrm{B}$. Only half of $\CoMP$ MSs can be simultaneously
active due to adopting coordinated scheduling. So, $\lambda_\mathrm{B}^\varphi$ is equal
to
$0.5\(1-\delta\)^2\lambda_\mathrm{B}+\(1-\delta\)\delta\lambda_\mathrm{B}=0.5\(1-\delta^2\)\lambda_\mathrm{B}$.
To sum up, the interferer densities can be quantified as follows:
\begin{align}
\lambda _{\mathrm{B}}^\psi &= \( \delta - 0.5 \delta^2 \) {\lambda _{\mathrm{B}}},   \label{E:densitypsi} \\
\lambda _{\mathrm{B}}^\varphi  &= 0.5 \left( {1 - \delta ^2 } \right){\lambda _{\mathrm{B}}}. \label{E:densityvarphi}
\end{align}
Then $\lt_\mathrm{U}^\varphi \left( s \right)$ can be expressed as:
\begin{align}
\label{E:ltvarphiul}
&\lt_\mathrm{U}^{\varphi}(s) = \exp \left( { - 2\pi \lambda _{\mathrm{B}}^\varphi \int_r^\infty  {\frac{{{\beta _{\mathrm{U}}}{r^\alpha }{y^{ - \alpha }}}}{{1 + {\beta _{\mathrm{U}}}{r^\alpha }{y^{ - \alpha }}}}ydy} } \right),
\end{align}
where $y$ denotes the distance from interfering UL-MSs to the typical BS. It is assumed
that the interfering UL-MSs are located at a distance larger than $r$. Using
\eqref{E:sucprobULCoMP2flex2}, \eqref{E:ltpsiul2}, and \eqref{E:ltvarphiul}, we can
evaluate the UL transmission success probability for $\CoMPtwoflex$.

\subsection{DL Success Probability in $\CoMPtwoflex$ Network}

Following the same lines as $p_\mathrm{U}^{\CoMPtwoflex}$, we can obtain the analytical
expression for the DL transmission success probability for $\CoMPtwoflex$
$p_\mathrm{D}^{\CoMPtwoflex}$ as follows:
\begin{align}\label{E:sucprobDLCoMP2flex}
p_\mathrm{D}^{\CoMPtwoflex} \approx \int_0^\infty  {\lt_\mathrm{D}^\psi \left( s \right)\lt_\mathrm{D}^\varphi \left( s \right)2\pi {\lambda _{\mathrm{B}}}r\exp \left( { - \pi {\lambda _{\mathrm{B}}}{r^2}} \right)dr},
\end{align}
where $s = \frac{\beta _\mathrm{D}r^\alpha}{P_\mathrm{B}}$ and the Laplace functionals of
the interference from BSs $\lt_\mathrm{D}^{\psi}\(s\)$ and MSs
$\lt_\mathrm{D}^{\varphi}\(s\)$ are
\begin{align}
\label{E:ltpsidl}
\lt_\mathrm{D}^{\psi}\(s\) &= \exp \left( { - 2\pi {\lambda _{\mathrm{B}}^\psi}\int_r^\infty  {\left( {\frac{{{\beta _{\mathrm{D}}}{r^\alpha }{x^{ - \alpha }}}}{{1 + {\beta _{\mathrm{D}}}{r^\alpha }{x^{ - \alpha }}}}} \right)xdx} } \right), \nonumber \\
\end{align}
\begin{align}
\label{E:ltvarphidl}
\lt_\mathrm{D}^{\varphi}\(s\) = \exp \left( { - 2\pi {\lambda _{\mathrm{B}}^\varphi}\int_r^\infty  {\frac{{\left( {{P_{\mathrm{M}}}/{P_{\mathrm{B}}}} \right){\beta _{\mathrm{D}}}{r^\alpha }{y^{ - \alpha }}}}{{1 + \left( {{P_{\mathrm{M}}}/{P_{\mathrm{B}}}} \right){\beta _{\mathrm{D}}}{r^\alpha }{y^{ - \alpha }}}}ydy} } \right).
\end{align}
The distance from the interfering DL-BSs to the typical DL-MS $x$ cannot be closer than
$r$ as shown in \cite{andrews2011tractable}. We further apply this distance restriction
to the distance from the interfering UL-MSs to the typical DL-MS $y$ to approximate. With
path-loss exponent $\alpha=4$, \eqref{E:ltpsidl} and \eqref{E:ltvarphidl} are further
simplified as follows:
\begin{align}
\label{E:ltpsidl2}
\lt_\mathrm{D}^{\psi}\(s\) = \exp \( - \pi \lambda_{\mathrm{B}}^{\psi}  \sqrt{\beta_{\mathrm{D}}} r^2 \arctan \( \sqrt{\beta_{\mathrm{D}}} \) \),
\end{align}
\begin{align}
\label{E:ltvarphidl2}
\lt_\mathrm{D}^{\varphi}\(s\) = \exp \( - \pi \lambda_{\mathrm{B}}^{\varphi}  \sqrt{ {\frac{P_\mathrm{M}}{P_\mathrm{B}}} \beta_{\mathrm{D}}} r^2 \arctan \( \sqrt{{\frac{P_\mathrm{M}}{P_\mathrm{B}}}\beta_{\mathrm{D}}} \) \).
\end{align}

\section{Performance Analysis}

\subsection{Success probability vs. SINR threshold}

We quantify the reliability performance of $\CoMPtwoflex$, where the analytical results
of \eqref{E:sucprobULCoMP2flex2} and \eqref{E:sucprobDLCoMP2flex} are compared with
numerical simulations using the greedy pairing algorithm. For comparison, we also present
the performances of $\CoMP$-only network, where BS cooperation happens only if two
serving BSs have the same traffic direction, and $\CoMPflex$-only network, where BS
cooperation happens only if two serving BSs have the cross traffic direction. We also
compare the greedy pairing algorithm with the Edmonds matching algorithm, in terms of
reliability and complexity. For the numerical simulations, the parameters used are shown
in Table~\ref{T:SimulationParameters}.

\begin{table}[tb]
	\caption{Simulation parameters.}
	\centering
			\begin{tabular}{ c l c }
		  	\hline
  			Parameter & Description & Simulation Setting\\
				\hline
				$S$ & Size of observation window & $150$ km \\
				$\lambda_\mathrm{B}$ & BS density & $0.02$ ${\text{BS}}/{\text{km}^2}$ \\
				$\sigma^2$ & Noise power at MS and BS & $-174$ dBm \\
				$\alpha$ & Path loss exponent & $4$ \\
				$\beta$ & SINR thresholds & $-15, -10,\ldots, 15$ dB \\
				$P_\mathrm{B}$ & BS transmission power & $40$ dBm \\
				$P_\mathrm{M}$ & MS transmission power & $20$ dBm \\
				$\delta$ & DL/(UL+DL)  & $0.5$ \\
				$W$ & System bandwidth & $1$ Hz \\
				$N$ & Simulation iterations & 10000 \\
				\hline
			\end{tabular}
	\label{T:SimulationParameters}
\end{table}

\begin{figure}[t]
	\centering
		\includegraphics[width=\linewidth]{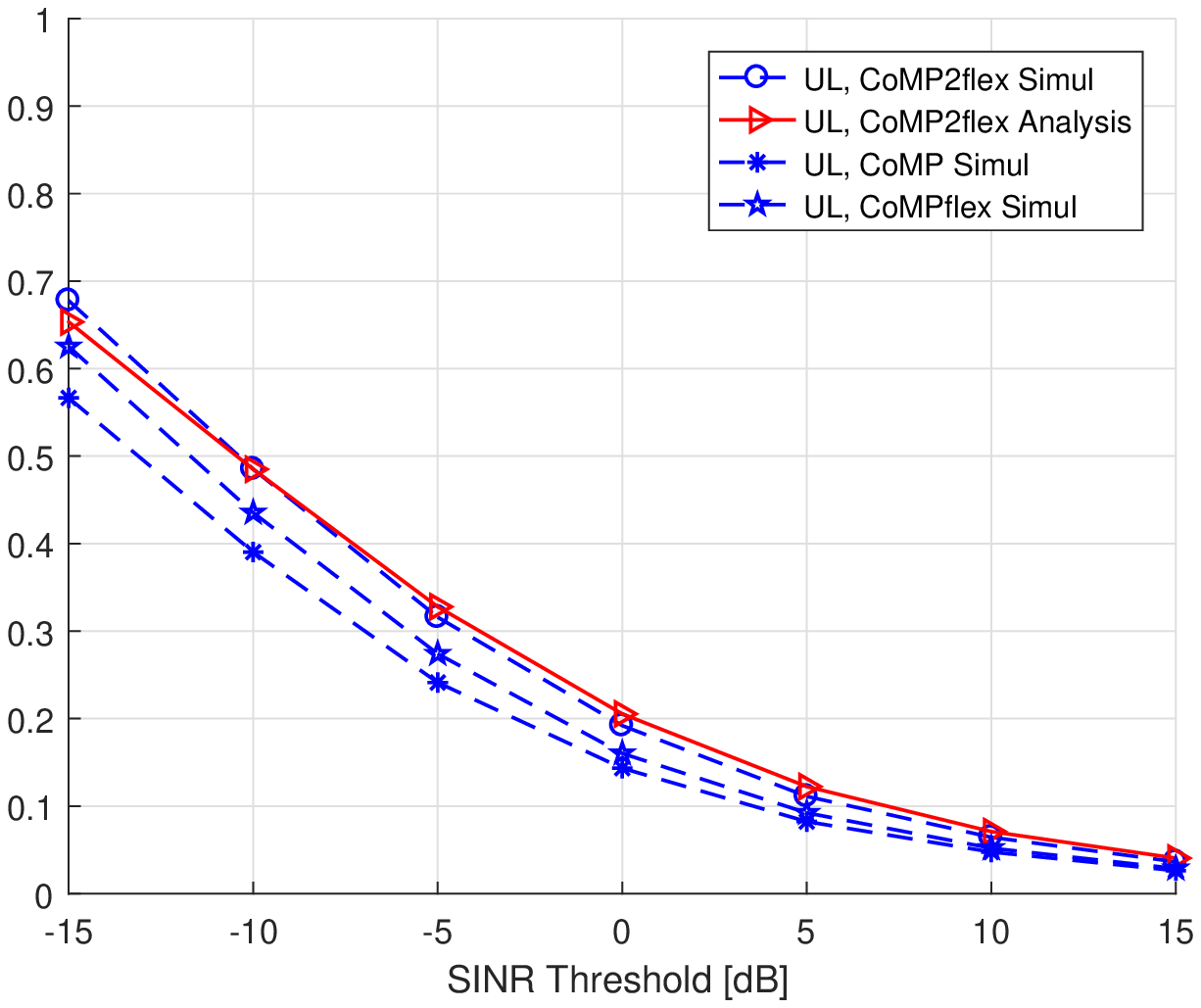}
	\caption{UL Success Probability vs. SINR threshold.}
	\label{F:sucprobul_vs_sir_thr}
\end{figure}

\begin{figure}[t]
	\centering
		\includegraphics[width=\linewidth]{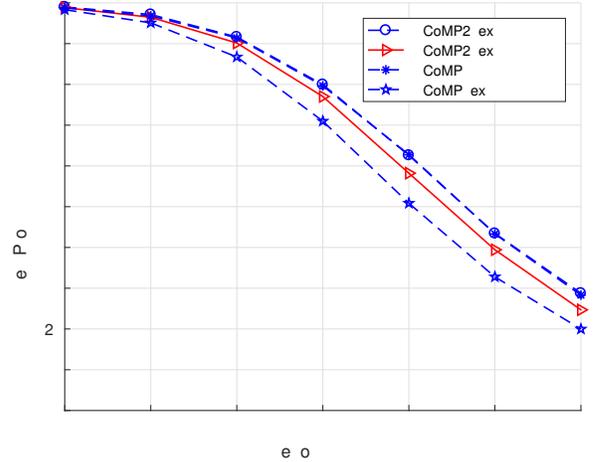}
	\caption{DL Success Probability vs. SINR threshold.}
	\label{F:sucprobdl_vs_sir_thr}
\end{figure}

The reliability performances in UL of $\CoMPtwoflex$, $\CoMP$-only and $\CoMPflex$-only
are shown in Fig.~\ref{F:sucprobul_vs_sir_thr}. We see that the analytical curve and
simulation follow the same trend, and that the performances of $\CoMP$-only and
$\CoMPflex$-only are lower than $\CoMPtwoflex$. This is because in $\CoMPtwoflex$, the
paired BSs use interference cancellation, yielding a better performance than
$\CoMP$-only, and protect the transmission by time sharing in case of both paired BSs
serving the same traffic direction, yielding a better performance than $\CoMPflex$-only
scheme. For the reliability in DL, shown in Fig.~\ref{F:sucprobdl_vs_sir_thr}, the
analytical and simulation curves again follow the same trend. Note that in this case,
since there is no interference cancellation between MSs (in contrast to UL),
$\CoMPtwoflex$ and $\CoMP$-only schemes have similar performance. But, still the better
performance is shown compared with $\CoMPflex$-only scheme because there is no protection
mechanism in $\CoMPflex$-only mode but $\CoMPtwoflex$ has it.

The small gap between the analytical and simulation curves can be explained from the
analytical model assuming PPP deployment of the MSs, while this is not the case in the
simulation, since each MS is deployed inside the cell of its BS as pointed in
\cite{kim2016reliable}.

\subsection{Comparison of the two matching algorithms}

\begin{figure}[t]
	\centering
		\includegraphics[width=\linewidth]{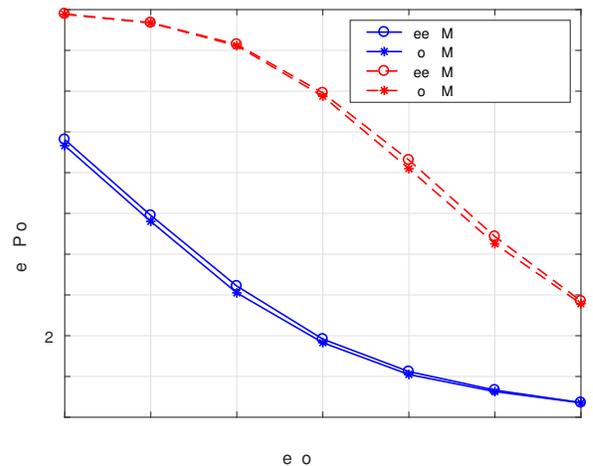}
	\caption{Success Probability in UL and DL vs. SINR threshold, comparing greedy and Edmonds matching, for $\CoMPtwoflex$.}
	\label{F:succULDLGreedyVsEdmonds}
\end{figure}

In this subsection, the performance of the two pairing algorithms, Greedy matching and
the Edmonds matching algorithm defined in Section~\ref{S:BS_Pairing_Algorithm}, is
compared in terms of reliability in UL and DL, in the $\CoMPtwoflex$ scheme. The
comparison is shown in Fig.~\ref{F:succULDLGreedyVsEdmonds}, where we see that for both
UL and DL, the greedy algorithm gives a slightly higher performance than Edmonds
matching. This is because how the BSs are paired in the two algorithms: in the greedy
one, each BS is paired to the \emph{closest} BS, while in the Edmonds matching, the goal
is to have as many BSs as possible paired, while at the same time minimizing the overall
distance.

We also compare the time in seconds required to run the two pairing algorithms, for a
range of BS densities, shown in Table~\ref{T:TimeMatching}. The computations were done on
a Quad-core 2.7 GHz computer with 16 GB RAM, running Windows 10 64 bit, and using Matlab
R2016b. We see that the greedy method is much faster than the considered implementation
of the Edmonds algorithm, which has a reported cubic complexity~\cite{EdmondsMatlab}.

\begin{table}[tb]
	\caption{Time comparison for the two matching algorithms.}
	\centering
			\begin{tabular}{ | c | l | c |}
		  	\hline
  			\textbf{BS density $\lambda_B$} & \textbf{Greedy alg.} & \textbf{Edmonds alg.} \\
				\hline
				0.0020 & 0.0005 & 0.0678 \\
				\hline
				0.0115 & 0.0027 & 2.2543 \\
				\hline
				0.0210 & 0.0050 & 6.8409 \\
				\hline
				0.0305 & 0.0074 & 15.9043 \\
				\hline
				0.0400 & 0.0100 & 24.9267 \\
				\hline
			\end{tabular}
	\label{T:TimeMatching}
\end{table}

From the comparison, we can conclude that using the greedy algorithm gives a performance
at least as good as the Edmonds one, an observation which validates
the method. Also, in the greedy algorithm, each BS needs only knowledge of its neighbors, compared to the Edmonds algorithm. 

\subsection{Throughput vs. DL Traffic Ratio}

\begin{figure}[t]
	\centering
		\includegraphics[width=\linewidth]{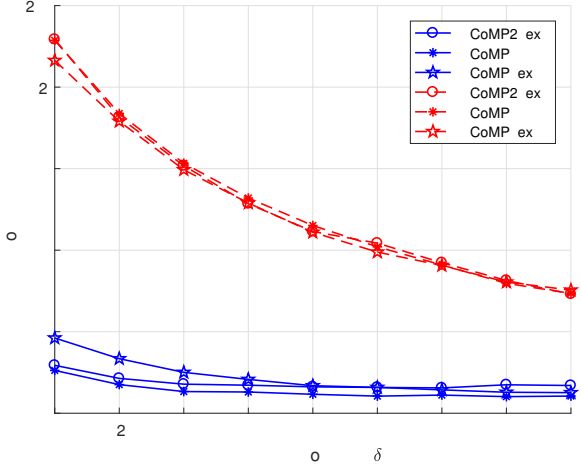}
	\caption{Throughput in UL and DL vs. DL Traffic Ratio.}
	\label{F:thptULDL}
\end{figure}

Even though we design the proposed $\CoMPtwoflex$ aiming increasing the reliability, a
throughput or spectral efficiency is important performance metric. The throughput of
$\CoMPtwoflex$ is compared with, $\CoMP$-only and $\CoMPflex$-only schemes in both UL and
DL traffic directions as a function of DL traffic ratio ($\delta$), in this subsection.
The throughput is measured in units of bits per seconds (bps) per Hz. In the cases of
$\CoMPflex$ and $\CoMP$-only schemes, since the BSs use time sharing (only on cooperating
BS in each $\CoMP$ pair is operating at a given time), the throughput is degraded by half
if the directions of traffic are same. The throughput performance of $\CoMPtwoflex$,
$\CoMPflex$-only and $\CoMP$-only is shown in Fig.~\ref{F:thptULDL}. The target threshold
is assumed as 10~dB. In this figure, we observe that in both UL and DL, all three schemes
follow a similar trend. Furthermore, they have a higher performance for UL-intensive
traffic (for $\delta \to 0$), which follows from the BS-BS interference cancellation.

For the DL direction (dashed curves), $\CoMPtwoflex$ and $\CoMP$-only are of similar
performance, owing to the time sharing. With low $\delta$, the performance of
$\CoMPflex$-only is the lowest.

For the UL direction (solid curves), $\CoMPflex$-only has higher performance than the
other two, from the BS-BS interference cancellation. With high $\delta$, the performance
of
$\CoMPtwoflex$ becomes superior.

\section{Concluding Remarks}

This paper has presented analytical expressions and numerical results on a reliability of
a BS cooperation scheme, $\CoMPtwoflex$, as well as comparing it with a variant of
$\CoMP$ using time sharing and its origin $\CoMPflex$. We can conclude that the impact of
BS-BS interference cancellation benefits of $\CoMPtwoflex$, suggesting it be suitable for
cooperation in cross directional traffic and its time sharing nature brings the benefits
in the cooperation of same directional traffic. We have also compared two pairing
algorithms, one where each BS needed only knowledge of its neighbors, and the other one,
the Edmonds matching algorithm, requiring full knowledge of the network. It was observed
that the performance attainable using these two algorithms are similar, which validates
the greedy algorithm used for the BS pairing. It could be interesting to investigate the
other $\CoMP$ modes to increase throughput performance of $\CoMPtwoflex$.

\section*{Acknowledgment}
This work has been in part supported by the European Research Council (ERC Consolidator
Grant Nr. 648382 WILLOW) within the Horizon 2020 Program and in part by Innovation Fund
Denmark, via the Virtuoso project.


\end{document}